\documentclass[%
aip,
rsi,
amsmath,amssymb,
reprint,
superscriptaddress,
]{revtex4-2}

\usepackage{graphicx}% Include figure files
\usepackage{dcolumn}% Align table columns on decimal point
\usepackage{bm}% bold math
\usepackage[utf8]{inputenc}
\usepackage[T1]{fontenc}
\usepackage{mathptmx}
\usepackage{etoolbox}

\makeatletter
\def\@email#1#2{%
 \endgroup
 \patchcmd{\titleblock@produce}
  {\frontmatter@RRAPformat}
  {\frontmatter@RRAPformat{\produce@RRAP{*#1\href{mailto:#2}{#2}}}\frontmatter@RRAPformat}
  {}{}
}%
\makeatother
\begin{document}

\title{Generation and Characterization of Surface-Attached Ultrathin Liquid Sheets for Grazing-Incidence X-ray Scattering}
\author{Yibo Wang} 
\affiliation{Stanford PULSE Institute, SLAC National Accelerator Laboratory, Menlo Park, California 94025, USA} 
\author{Daniel P. DePonte} 
\affiliation{Linac Coherent Light Source, SLAC National Accelerator Laboratory, Menlo Park, California 94025, USA} 
\author{Adi Natan} 
\email[Corresponding author: ]{natan@stanford.edu} 
\affiliation{Stanford PULSE Institute, SLAC National Accelerator Laboratory, Menlo Park, California 94025, USA}
\date{\today}

\begin{abstract}
Capturing ultrafast structural dynamics at solid-liquid interfaces is key to understanding adsorption, desorption, diffusion, and aggregation processes in catalysis and interfacial chemical reactions. Hard-X-ray scattering in grazing-incidence geometry can, in principle, access interfacial structural changes with angstrom-scale structural sensitivity and ultrafast temporal resolution.
However, the long optical paths of the optical pump and hard-X-ray pulses inside the liquid sample pose significant challenges to the temporal resolution, signal-to-noise ratio, and overall stability of such an experimental scheme.
Here, we report a method for creating and characterizing ultrathin surface-attached free-flowing liquid sheets, whose submicrometer thickness enables ultrafast temporal resolution and reduces the bulk-liquid scattering contribution.
The impinging-jet geometry produces stable micrometer-scale sheets whose morphology depends systematically on incidence angle, jet velocity, and capillary diameter.
Gas-assisted shaping using a second capillary further narrows and thins the sheet, producing an extended ultrathin region and reducing the measured minimum thickness below 500~nm for acetonitrile.
The resulting platform provides a reproducible, continuously flowing, surface-attached liquid geometry for grazing-incidence scattering experiments.
\end{abstract}
\maketitle
 \section{Introduction}

The ultrafast dynamics at the interface between solids and liquids are of substantial importance for understanding catalytic activities and chemical processes.
In the past, such interfaces have been investigated in static settings  \cite{tsuyumoto2000ultrafast,magnussen2019toward,harmon2020validating}.
Scattering methods provide a pathway for in situ, direct measurements of the structural dynamics that occur at a solid-liquid interface  \cite{fenter2002xray,wang1992situ,magnussen2024situ}.
Recent ultrafast hard-X-ray scattering work has shown that weak  optical-pump/X-ray-probe difference signals can be interpreted in terms of atomic-scale structural dynamics, anisotropic scattering components, and real-space pair-density changes, provided that the sample geometry, background, and temporal response are well controlled~\cite{natan2021resolving,natan2023realspace,powersriggs2024deformational,schori2025realspace,hopper2026coherent,stamm2026real}. Extending this capability to solid-liquid interfaces requires a continuously refreshed liquid layer that is thin enough to suppress bulk background and optical group-delay smearing while remaining laterally stable over the grazing-incidence X-ray footprint. The central experimental challenge is therefore to prepare an ultrathin, surface-attached liquid film that satisfies these constraints.
\par
Recent developments in the design and manufacturing of microfluidics have enabled the generation and characterization of free-flowing liquid sheet jets with sub-micrometer thicknesses \cite{galinis2017micrometer,koralek2018generation,ha2018device,menzi2020generation,crissman2022sub}.
The ultrathin films of liquid generated from such apparatuses enable the direct observation of the ultrafast dynamics of liquids and solutions, using ultrafast X-ray spectroscopies \cite{ekimova2015liquid,smith2020femtosecond,loh2020observation} and ultrafast electron diffraction \cite{nunes2020liquid,yang2021direct}. 
In contrast to free-standing ultrathin liquid sheets, the generation and characterization of surface-attached ultrathin liquid sheets remain underexplored.

Static surface-attached wet films have been reported to provide nanometer-scale thicknesses, which are ideal for static investigations of solid-liquid surfaces using X-ray photoelectron spectroscopy approaches \cite{axnanda2015using,novotny2020probing,martins2021near}.
However, the lack of continuous renewal makes these methods prone to pump-induced heating, photochemical accumulation, contamination, and X-ray-induced effects.
They are therefore poorly suited to high-repetition-rate optical-pump/X-ray-probe measurements and to condition-modulation experiments such as UV-on/UV-off cycling.

Oblique impingement of circular liquid jets on solid surfaces is known to produce non-axisymmetric spreading sheets and hydraulic-jump boundaries, with the stagnation point shifted relative to the geometric center of the projected impingement footprint~\cite{kate2007hydraulic,mishra2020experimental}. Prior work has shown that the spreading profile depends on incidence angle, jet velocity, nozzle size, surface tension, and surface wetting~\cite{kibar2010experimental,kibar2018spreading,li2021effect}. In hydrophobic and superhydrophobic cases, related flows can transition among spreading, braiding, reflection, rebound, and splashing or droplet-emission regimes~\cite{celestini2010water,kibar2010experimental}. We use this fluid-mechanical framework not to study rebound or splashing, but to identify operating conditions that maintain a stable, continuously renewed, surface-attached ultrathin liquid sheet on a solid substrate.

\begin{figure}
    \centering
    \includegraphics[width=\columnwidth]{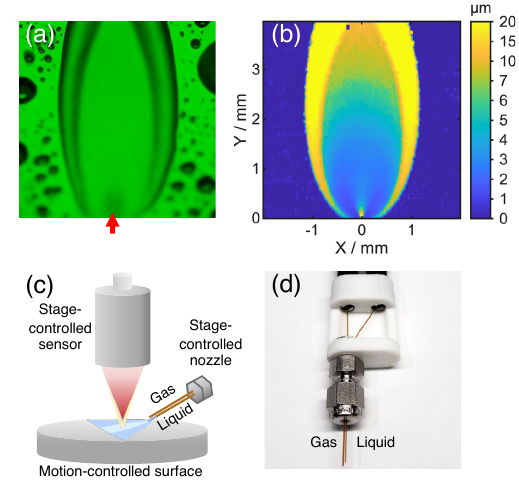}
   \caption{
Experimental concept and measurement geometry.
(a) Water sheet formed by oblique impingement of a liquid microjet on a flat solid substrate. Small ejected droplets are visible outside the intended region of interest under this representative high-impact condition.
(b) Reconstructed thickness map of a representative water sheet; droplets and disconnected reflections were masked during reconstruction. The faint leaf-shaped ridge inside the outer rim is an inner capillary rim associated with lateral flow focusing upstream of the hydraulic jump.
(c) Schematic of the confocal displacement sensor, dual-capillary nozzle, and substrate geometry. The sensor is mounted on motorized X--Y stages for raster scanning. The nozzle assembly is independently positioned using X, Y, and Z translation stages and one rotary stage. The solid substrate is mounted on a liquid catcher supported by a tilt--yaw stage, allowing the substrate plane to be aligned with the rasterized scan pattern.
(d) Photograph of the dual-capillary nozzle used for gas-assisted shaping.
}
 
    \label{fig1}
\end{figure}

Here we report an experimental platform for generating and characterizing continuously renewed, surface-attached ultrathin liquid sheets. The concept is summarized in Fig.~\ref{fig1}. A liquid microjet impinges obliquely on a flat solid substrate, redirecting the incident momentum into a surface-attached liquid sheet that spreads outward from the impingement zone and terminates at a hydraulic jump, as shown in Fig.~\ref{fig1}(a). Small ejected droplets are visible in this representative high-impact condition; these droplets lie outside the intended X-ray interaction region and are avoided in the optimized operating window. The corresponding reconstructed thickness map in Fig.~\ref{fig1}(b) illustrates how raster-scanned chromatic confocal displacement sensing converts the flowing free-surface geometry into a quantitative film-thickness map. The measurement geometry is shown schematically in Fig.~\ref{fig1}(c), where the confocal sensor, liquid-delivery nozzle, and solid substrate are independently positioned to control the incidence angle, nozzle--surface distance, and scan plane. Finally, Fig.~\ref{fig1}(d) shows the dual-capillary nozzle that was developed and used in the gas-assisted configuration, in which an adjacent gas jet reshapes and thins the surface-attached sheet. Using this platform, we map the dependence of water-sheet morphology on incidence angle, jet velocity, and nozzle diameter, and then show that gas-assisted acetonitrile sheets can reach the 250~nm instrumental readout limit of the displacement sensor while providing a millimeter-scale sub-500-nm region suitable for grazing-incidence hard-X-ray scattering.

\section{Experimental setup}
\label{sec:setup}

The experimental setup comprises a liquid circulation loop and an enclosed, ventilated sample environment that integrates an actuated measurement assembly.
The enclosure maintains continuous air exchange and a gentle negative pressure to remove hazardous solvent vapor, ensuring safe operation during measurements.
The following subsections focus on the liquid circulation system, the chromatic confocal measurement setup, and the adjustable interaction region that defines the sensor, nozzle, and substrate geometry.

\subsection{Liquid circulation system}

The liquid was stored in a sealed reservoir and continuously circulated through the system during data acquisition.
The water measurements used $18.2~\mathrm{M\Omega\,cm}$ deionized water, while the acetonitrile measurements used 99.9\% HPLC-grade acetonitrile.
Before use, the sample reservoir was filled with solvent, sealed, placed in an ultrasonic bath, and connected to a vacuum pump to remove dissolved gases that could nucleate bubbles or destabilize the jet.
The vacuum-sonication time was 1~h for water and 20~min for acetonitrile.
\par
After degassing, the sample was connected to the circulation loop, and the flow was allowed to reach steady operation before measurements were performed.
A high-pressure liquid chromatography (HPLC) pump (Shimadzu LC-20AP) drove the liquid through a high-pressure line to a fused-silica capillary nozzle with an outer diameter of 375~\textmu m and a variable inner diameter depending on the desired jet size.
The mean liquid-jet velocity is calculated from the pump flow rate $Q$ and capillary inner diameter $d_c$ as $V_\ell=4Q/(\pi d_c^2)$.
An intake filter and an inline filter minimize the risk of clogging.
Upon impingement on the solid substrate, the liquid spreads into a thin, surface-attached sheet and is subsequently collected by a catcher that channels the spent liquid back to the reservoir, allowing continuous recirculation under steady-state conditions.

\subsection{Measurement system}

A motion-controlled confocal displacement sensor (Keyence CL-S015) is employed to characterize the thickness and topography of the liquid sheet.
The sensor operates on the principles of chromatic confocal microscopy.
A chromatic lens system disperses a broadband white-light source, spatially separating the wavelengths along the optical axis.
Each wavelength is focused at a unique distance from the lens system, and the reflected spectrum from the sample surface is collected.
The wavelength of maximum reflected intensity is then converted into displacement through a calibrated wavelength--position relationship, providing high-accuracy, non-contact distance measurement~\cite{li2024high,li2021thickness,liu2024new}.
The displacement measurement has a readout resolution of 250~nm, while the lateral spatial resolution is limited by the 15~\textmu m optical spot size of the sensor.
The measurement repetition rate can be adjusted between 1~kHz and 10~kHz, enabling rapid acquisition of topographical maps and temporal stability measurements under various operating conditions.

The sensor head is mounted on a pair of motorized X--Y translation stages (Newport) and a manual Z stage, providing lateral scanning and focus control for automated mapping of the liquid sheet over designated areas.
The manual Z stage is used to zero the sensor and maintain the optimal working distance during measurements.
The entire measurement setup is secured on an optical table to minimize vibrations transmitted from the motion stages.
To reconstruct the three-dimensional profile of the liquid sheet, the sensor records displacement data over a predefined raster-scan pattern.
The spatially resolved displacement values are compiled into a height map, such as the one shown in Fig.~\ref{fig1}(b).
This configuration enables reproducible and spatially resolved profiling of the local film thickness with submicrometer vertical precision.

\subsection{Liquid-sheet generation and measurement geometry}

Figure~\ref{fig1}(c) depicts the interaction region where the liquid sheet is generated and characterized.
The confocal displacement sensor is positioned above the interaction zone, with the solid substrate below and the liquid nozzle on the right-hand side.
The solid surface is a protected silver mirror (Thorlabs, PF20-03-P01) mounted on top of the liquid catcher, which is affixed to a manual yaw and tilt stage.
This stage enables fine adjustment of the surface orientation to ensure parallel alignment between the mirror plane and the scan grid of the confocal sensor.
This alignment ensures a uniform scan background, which is critical for accurate reconstruction of the sheet topography and local film-thickness distribution.

The liquid-jet assembly provides four degrees of freedom: translation along the X, Y, and Z axes, together with angular adjustment, allowing precise control of the jet incidence on the substrate.
Throughout this work, the incidence angle is defined as the angle between the liquid-jet axis and the substrate plane.
The nozzle--surface distance was kept below 8~mm, short enough to avoid jet breakup before impingement.

For the dual-capillary configuration shown in Fig.~\ref{fig1}(d), a 3D-printed holder aligns two glass capillaries within a dual-lumen channel, positioning their parallel exits approximately 300~\textmu m apart.
The lower capillary delivers the liquid jet, while the upper capillary supplies the assisting gas jet. 
The dual-capillary holder provides a nominally co-propagating gas--liquid geometry at the nozzle exit, while the final gas footprint on the sheet is set by small translations and rotations of the nozzle assembly. 
By tuning the gas backing pressure and the distance between the gas jet and the liquid sheet, the lateral spread and thickness distribution of the liquid sheet can be further optimized.
Alternatively, the liquid and gas nozzles can be mounted on separate opto-mechanical holders to access a wider range of relative angles between the gas and liquid jets, although this configuration increases the minimum achievable separation between the two nozzle tips. 
These configurations allow the momentum transfer from the assisting gas jet to be directed either nearly along the liquid-jet projection, as in the acetonitrile case, or at a larger angular offset, as in the water case.
The gas jet provides an additional means of shaping the sheet profile, consistent with previous gas-assisted focusing and gas-accelerated liquid-sheet approaches~\cite{ganan1998generation,deponte2008gas,belsak2021computational,zahoor2018influence,kovacic2026emergence}. The overall architecture builds on liquid-delivery strategies developed for compact liquid injection and high-repetition-rate liquid-sheet delivery at X-ray free-electron lasers~\cite{knovska2020ultracompact,hoffman2022microfluidic,konold2023printed}.

\subsection{Operations}

Before the parameter sweeps, the liquid sample was prepared, and the measurement procedure was validated under fixed operating conditions.
The displacement readout and scan trajectory were checked using a standard 3.05-mm-diameter, 20~\textmu m-thick TEM grid, whose reconstructed height and grid geometry agreed with the known specifications.
The mirror plane was then aligned to the lateral scan plane, and the sensor height was adjusted to the optimal working distance near the center of the chromatic detection range.

During sheet profiling, the chromatic confocal sensor recorded the reflected signal from the liquid--air interface in the sheet and from the exposed substrate in regions outside the sheet.
The local film thickness was reconstructed as the height of the liquid--air interface relative to the local substrate plane.
In the standard scanning procedure, a rasterized line-scan protocol was used to retrieve the topographical maps, in which the fast axis was scanned continuously across the sheet while the second translation stage was incremented after each line.
The scanned area was typically on the order of several millimeters, with a slow-axis step size of 20--50~\textmu m.
The fast-axis sampling interval was determined by the stage speed and the sensor acquisition rate; for example, at a line-scan speed of 2~mm/s and an acquisition rate of 1~kHz, the raw sampling interval along the fast axis was approximately 2~\textmu m.
This scan speed reduces stage-induced vibration while oversampling the line scans, allowing the data to be downsampled to a spatial resolution comparable to the optical spot size without artifacts.
A 0.5~s wait time was used between stage movements to allow the stages to settle.
For sparse stability measurements, the sensor was held fixed at each grid point for 10~s, and the standard deviation of the displacement trace was used to quantify local temporal fluctuations.
\par
In the data-processing workflow, the raw displacement traces were converted to spatial maps using the recorded timing and stage-position data.
Spurious points from droplets, loss of signal, or reflections outside the liquid sheet were rejected by thresholding and by masking regions outside the detected sheet boundaries.
 
Residual substrate offset and tilt were removed by fitting the exposed mirror regions outside the sheet. 
Line-to-line height offsets associated with translation-stage hysteresis during raster scanning were corrected, when possible, using exposed-substrate regions outside the liquid sheet as the zero-thickness reference.
This correction removes most row-wise offsets, but residual scan-line modulations can remain when the liquid sheet fills the scan width and the exposed substrate cannot be sampled on both sides of the sheet.
This limitation is most relevant for the gas-assisted acetonitrile map, where the broad sheet prevented a reliable line-by-line substrate correction. Small residual modulations in the longitudinal lineout should therefore not be interpreted as physical thickness oscillations.
The processed data were smoothed along the scan direction using a 20--50~\textmu m moving window and interpolated onto a uniform grid for plotting and quantitative comparison. Because the displacement readout resolution is 250~nm, minimum measured values near this level should be interpreted as instrument-limited. 
\section{Results}
 
The reconstructed three-dimensional profile of the surface-attached water sheet is shown in Fig.~\ref{fig1}(b).
This result confirms that the impinging-jet configuration can reliably produce a free-flowing, surface-attached liquid sheet.
Near the centerline of the sheet lies the impingement zone, where the liquid jet impacts the surface.
As the liquid propagates radially outward, friction at the solid-liquid interface gradually slows the flow, leading to a steady increase in sheet thickness with distance from the point of impact \cite{watson1964radial,liu1993hydraulic}.
When the flow velocity falls below the condition required to sustain the rapidly spreading thin film, the film undergoes a hydraulic jump, producing a sharp rise in height that defines the outer rim of the sheet \cite{craik1981circular,liu1993hydraulic,bhagat2018origin}.
In addition to the outer hydraulic-jump rim, the thickness maps show a weaker leaf-shaped ridge upstream of the jump. 
We assign this feature to an inner capillary rim formed by lateral redistribution of the obliquely impinging flow, rather than to a second hydraulic jump. 
Similar rim and inner-contour structures have been reported for inclined impinging jets, where inertia drives the initial spreading while surface tension collects the flow into prominent side rims before recombination or jump formation~\cite{kibar2010experimental,li2021effect,kate2007hydraulic}.
The regions of interest are the two thin regions immediately adjacent to the impingement zone on either side of the centerline, where the sheet thickness is minimized.
\par
For hard-X-ray scattering experiments performed in grazing-incidence geometry, the relevant optimization criteria are the minimum and average film thickness within the illuminated region, the temporal stability of that region, and the continuous sheet length available along the projected X-ray footprint. The useful region is therefore not defined by a universal thickness threshold, but by the selected grazing angle, beam size, solvent path length, and acceptable pump-probe temporal smearing. Accurate characterization of the thickness, temperature, and stability of thin liquid targets is therefore essential for interpreting ultrafast X-ray and electron scattering measurements \cite{buttersack2023imaging,chang2022temperature}.
Within the tested range, the nozzle--surface distance does not significantly affect the surface morphology of the liquid sheet, provided that the distance remains small enough to prevent breakup of the impinging jet.
In addition, the liquid sheet must maintain stable flow conditions over an extended period of operation \cite{barnard2022delivery,hoffman2022microfluidic}.
The reproducibility of the reconstructed profiles demonstrates the long-term stability of the sheets under these conditions.
We next examine the short-time-scale temporal stability of the sheets, which is critical for maintaining a consistent interaction geometry in pump-probe experiments.

\subsection{Sheet stability}

\begin{figure}
    \centering
    \includegraphics[width=\columnwidth]{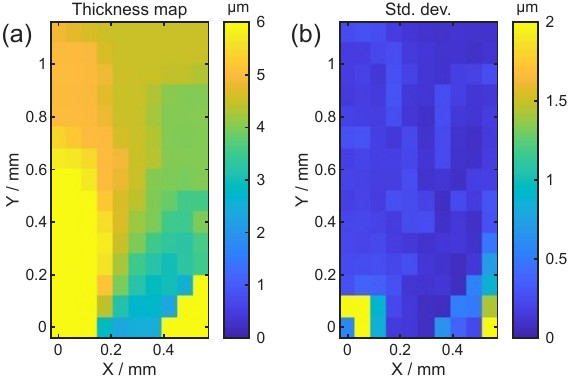}
    \caption{
    (a) Reconstructed thickness map of a representative surface-attached liquid sheet (water on protected silver mirror), obtained from a sparse grid scan, shown zoomed into the area surrounding the region of interest. 
    (b) Corresponding stability map showing the standard deviation of the measured thickness, obtained from the same scan pattern in which the confocal sensor hovered at each point for 10~s at a 1~kHz sampling rate.
Within the region of the flowing sheet that excludes the impingement zone and the hydraulic-jump rim, the median measured temporal standard deviation was 176~nm, comparable to the 250~nm readout resolution of the confocal sensor. }
    \label{fig3}
\end{figure}

Figure~\ref{fig3}(a) shows a representative thickness map of a surface-attached liquid sheet, while Fig.~\ref{fig3}(b) presents the corresponding stability map, expressed as the standard deviation of the measured thickness.
The liquid sheet is generated using a capillary of 75~\textmu m inner diameter with a jet velocity of 34 m/s.
The stability was evaluated using a sparse grid measurement, in which the confocal sensor hovered over each grid point for 10 s at a 1~kHz data readout rate to collect a sufficient number of displacement readings for statistical analysis.
The resulting standard deviation quantifies the temporal fluctuations in the local film thickness.
Across the region of interest, the measured temporal fluctuations are close to the displacement readout floor of the instrument. The median measured standard deviation was 176~nm, comparable to the 250~nm readout resolution of the confocal sensor, indicating that the local film thickness is stable at the instrument-limited level under steady-state operation.

\subsection{Angle dependence}

\begin{figure*}
    \centering
    \includegraphics[width=5.3 in]{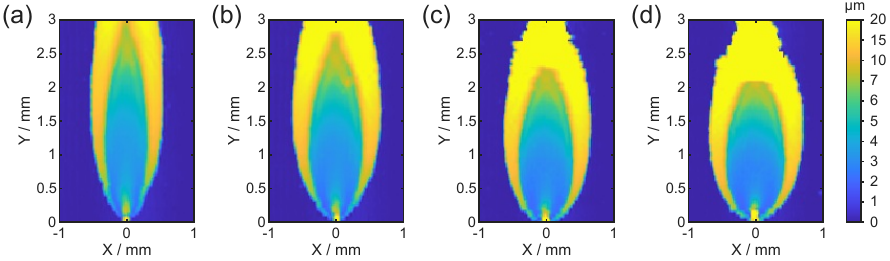}
    \caption{
    Reconstructed spread profiles of water on a silver mirror at incidence angles of (a) 15\textdegree, (b) 20\textdegree, (c) 25\textdegree, and (d) 30\textdegree. 
    A shallower incidence angle produces an elongated sheet with a larger minimum thickness and a smaller total area. 
    Increasing the angle improves spreading and expands the thin-sheet region, but excessive impact at high angles contributes to droplet formation near the impingement zone, limiting stable operation near 30\textdegree{} under the tested conditions.}
    \label{fig4}
\end{figure*}

Figure~\ref{fig4} shows the reconstructed spread profiles of water on a silver mirror substrate at incidence angles of 15\textdegree, 20\textdegree, 25\textdegree, and 30\textdegree{} in panels (a)--(d), respectively.
Here, the incidence angle is defined as the angle between the jet axis and the surface plane.
The liquid sheets are generated from a capillary of 75~\textmu m inner diameter at a jet speed of 45.4~m/s. 
At shallower incidence, the sheet is elongated along the jet-projection direction, the total wetted area is smaller, and the central reflected-flow feature downstream of the impingement zone is more pronounced.
As the incidence angle increases, the flow spreads more efficiently in the transverse direction: the thin-sheet area expands and the minimum measured film thickness decreases~\cite{kate2007hydraulic,kibar2010experimental,li2021effect}.
However, for a fixed jet velocity, increasing the incidence angle also leads to a greater impact force at the interface.
Beyond a threshold, visible ejected droplets appear near the impingement zone and around the sheet periphery, as shown in Fig.~\ref{fig1}(a).
Such droplets should be avoided for grazing-incidence X-ray measurements because they introduce an uncontrolled scattering background and may intercept the incident or scattered beam. 
The optimal incidence angle depends on the jet velocity. Under the present experimental conditions, however, angles near 30\textdegree{} provide the largest stable thin-sheet region before droplet emission becomes significant.

\subsection{Velocity dependence}

\begin{figure*}
    \centering
    \includegraphics[width=5.6 in]{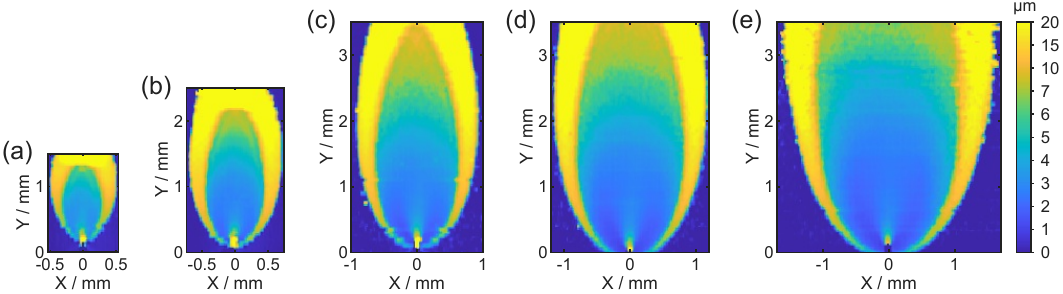}
    \caption{
Reconstructed profiles of water sheets generated from a 75~\textmu m inner-diameter capillary at jet velocities of (a) 11.3, (b) 22.7, (c) 34, (d) 45.3, and (e) 56.7~$\mathrm{m/s}$, all at a 30\textdegree{} incidence angle. 
Increasing jet velocity enhances outward propagation before the onset of hydraulic jumps, expanding the total sheet size. 
The minimum measured film thickness decreases with velocity, but the reduction becomes marginal above approximately 34~$\mathrm{m/s}$. At still higher speeds, visible droplet emission destabilizes the flow.}
    \label{fig5}
\end{figure*}

Figure~\ref{fig5} shows the reconstructed profiles of water sheets at jet velocities of 11.3, 22.7, 34, 45.3, and 56.7~$\mathrm{m/s}$ in panels (a)--(e), respectively, all taken at a 30\textdegree{} incidence angle. The jet velocity $V_\ell$, defined in Sec.~\ref{sec:setup}, was controlled by the HPLC pump flow rate. All measurements in Fig.~\ref{fig5} used a 75~\textmu m inner-diameter capillary. For water sheets on a silver mirror generated from this capillary without gas assistance, the minimum measured film thickness is approximately 1.7~\textmu m, and the region of interest extends over a characteristic length of approximately 200~\textmu m.
Increasing the jet velocity enhances the initial speed at the impingement zone, resulting in greater outward propagation distance before the onset of the hydraulic jumps \cite{watson1964radial,liu1993hydraulic,kuraan2017heat}.
Consequently, the total sheet size expands with velocity. 
However, the reduction in minimum thickness does not scale linearly with velocity; it decreases rapidly at low velocities, but the reduction becomes marginal above approximately 34~$\mathrm{m/s}$, suggesting a balance between inertial thinning, viscous dissipation, and capillary restoring forces.
At higher speeds, the increased impact force promotes visible droplet emission at the interface \cite{xu2005drop}.
To avoid this effect, the maximum operational jet velocity is limited to below 60~$\mathrm{m/s}$ under the tested conditions.
A quantitative comparison of the corresponding Reynolds, Weber, Capillary, and Froude numbers is given in Sec.~III.E.

\subsection{Jet size dependence}

\begin{figure}
    \centering
    \includegraphics[width=3.2 in]{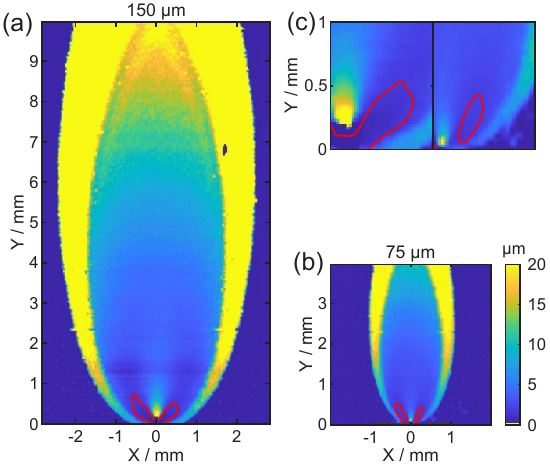}
    \caption{
    Reconstructed spread profiles of water on a silver mirror using capillary nozzles with inner diameters of (a) 150~\textmu m and (b) 75~\textmu m, both operated at a jet velocity of 34~$\mathrm{m/s}$.
    The red contour lines mark a 2~\textmu m reference thickness used to compare the relative size of the thinnest continuous regions.
    The larger nozzle produces a proportionally larger sheet while maintaining a comparable minimum thickness. 
    (c) Zoomed comparison of the regions enclosed by the 2~\textmu m contours for the 150~\textmu m capillary on the left and the 75~\textmu m capillary on the right.    }
    \label{fig6}
\end{figure}

The reconstructed spread profiles of water on a silver mirror for two capillary nozzles with inner diameters of 150~\textmu m and 75~\textmu m are displayed in Fig.~\ref{fig6}(a) and (b), respectively, both operated at a jet velocity of 34 $\mathrm{m/s}$.
The larger-diameter jet produces a proportionally larger sheet, consistent with the increased volumetric flow rate and impingement zone size \cite{watson1964radial,olsson1966radial}.
The minimum sheet thickness remains comparable between the two cases, indicating that the local film thinning near the impingement zone is primarily governed by the jet velocity and distance to the impingement zone, rather than the overall flow volume.
Despite the proportional increase in total sheet size, the area where the film thickness remains below 2~\textmu m does not change proportionally with jet diameter.
The thicker peripheral regions limit the effective usable area for ultrathin film applications. The dimensionless scaling analysis below further clarifies why increasing the nozzle diameter expands the sheet area without producing a proportional reduction in minimum thickness.
The red contours in Fig.~\ref{fig6}(c) also identify the thin side-lobe regions adjacent to the impingement zone. 
These regions are flatter than the central reflected-jet feature and therefore provide the most natural target for subsequent gas-assisted shaping.
 
\subsection{Dimensionless operating range}
Although the present measurements are primarily empirical thickness maps, the dimensionless operating range helps place the observed trends in the context of impinging-jet flows and clarifies which forces are expected to control the initial sheet thinning. We use these quantities only to characterize the tested regime, not as a predictive model for the non-axisymmetric hydraulic-jump shape.
    Because the free-jet diameter was not independently measured and is expected to be close to the capillary inner diameter under these flow conditions, we use the capillary inner diameter $d_c$ as the characteristic length. The relevant dimensionless groups are
\begin{equation*}
Re{=}\frac{\rho V_\ell d_c}{\mu}, \qquad
We{=}\frac{\rho V_\ell^2 d_c}{\sigma}, \qquad
Ca{=}\frac{\mu V_\ell}{\sigma}, \qquad
Fr{=}\frac{V_\ell}{\sqrt{g d_c}},
\end{equation*}
where $V_\ell$ is the mean liquid-jet velocity, $\rho$ is the liquid density, $\mu$ is the dynamic viscosity, $\sigma$ is the surface tension, and $g$ is the gravitational acceleration.
For the water velocity series using a 75~\textmu m capillary, the tested range of $11.3$--$56.7~\mathrm{m\,s^{-1}}$ corresponds to $Re\approx9.5\times10^2$--$4.8\times10^3$, $We\approx1.3\times10^2$--$3.3\times10^3$, $Ca\approx0.14$--$0.70$, and $Fr\approx4.2\times10^2$--$2.1\times10^3$.
The large Froude numbers indicate that gravity is not expected to control the initial spreading and thinning near the impingement zone, although it can still influence the downstream drainage and hydraulic-jump region.
Under these conditions, the reduction in minimum measured film thickness becomes marginal above approximately 34~$\mathrm{m\,s^{-1}}$, consistent with a regime in which inertial thinning is increasingly balanced by viscous dissipation, capillary restoring forces, the evolving hydraulic-jump boundary, and the onset of visible droplet emission.

\subsection{Gas-assisted thinning of liquid sheets}

\begin{figure*}
    \centering
    \includegraphics[width=5.0 in]{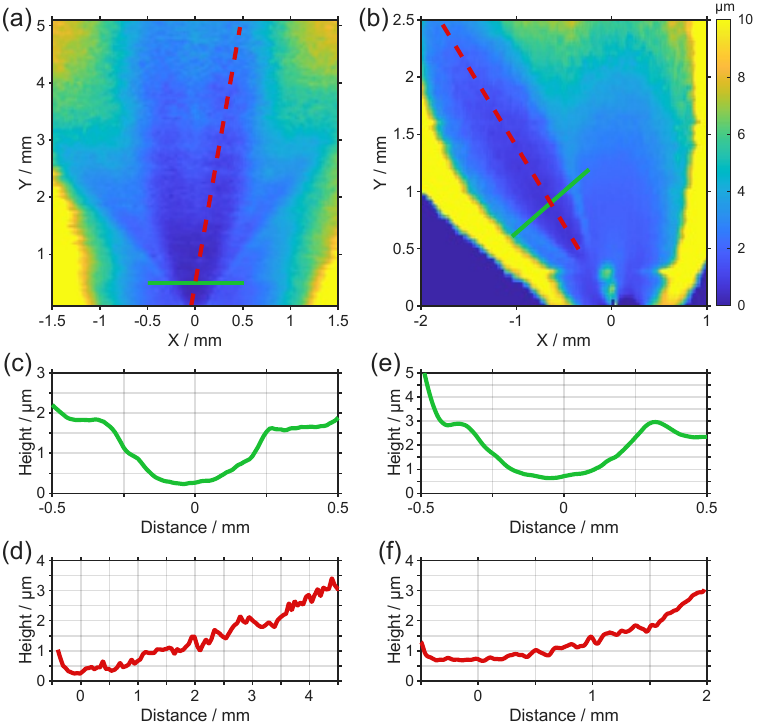}
   \caption{
Thickness profiles of liquid sheets shaped by assisting gas jets.
(a) Reconstructed thickness profile of an acetonitrile sheet shaped by a gas jet that remains approximately collinear with the liquid jet but is laterally displaced onto a thin side-lobe region of the sheet.
(b) Reconstructed thickness profile of a water sheet shaped using a larger gas-liquid angular offset to direct the gas-driven forcing onto a similarly flat side-lobe region.
In both panels, the dashed red line indicates the nominal gas-jet trajectory, while the liquid-jet projection is along the $Y$ direction in the plotting convention used here.
(c) and (d) Color-coded X- and Y-direction lineouts extracted from the acetonitrile map in (a).
The Y-direction lineout is slightly tilted to follow the thinned channel created by the gas jet.
(e) and (f) Corresponding X- and Y-direction lineouts extracted from the water map in (b).
}
    \label{fig8}
\end{figure*}

After optimizing the liquid-sheet geometry through capillary size, incidence angle, and jet velocity, we introduced a dual-capillary nozzle, displayed in Fig.~\ref{fig1}(d), to provide an additional gas-assisted thinning mechanism.
The resulting reconstructed profiles of liquid sheets of acetonitrile and water are shown in Fig.~\ref{fig8}(a) and Fig.~\ref{fig8}(b), respectively.
Both measurements were performed at a 25\textdegree{} impingement angle using a 75~\textmu m liquid capillary and a 150~\textmu m gas capillary.
The liquid-jet velocity was 34~$\mathrm{m\,s^{-1}}$.
Compressed air was supplied to the gas nozzle at an upstream pressure of approximately 25~psi without active flow regulation, under these conditions, the gas flow was limited by the throughput of the capillary.
Before liquid-sheet characterization, the free gas jet was measured in the same nozzle configuration using an external gas-flow velocimeter positioned 1~cm downstream of the gas-nozzle exit, giving a reference gas speed of approximately 27.4~$\mathrm{m\,s^{-1}}$.

During these measurements, the gas nozzle was lowered to the closest position that maintained sheet continuity.
The gas alignment was optimized to apply gas-driven forcing to the smoothest, already thin part of the liquid sheet rather than to the geometric centerline of the liquid jet.
As discussed above, the baseline maps show that the most useful thin regions are the side lobes adjacent to the impingement zone, while the central direction contains a reflected-jet feature and stronger curvature.
For acetonitrile, the liquid and gas trajectories remained approximately collinear, but the dual-capillary assembly was slightly rotated and translated so that the gas footprint was displaced laterally onto the thin side-lobe region.
For water, a larger gas--liquid angular offset was required to keep the gas-driven forcing on a similarly flat side-lobe region while preserving a continuous sheet.
In both cases, the alignment was chosen to maximize the length and flatness of the gas-thinned channel.

\par
 
The introduction of the gas jet substantially reshapes the liquid sheet and reduces the local film thickness.
For acetonitrile, the measured minimum reached the 250~nm readout limit of the confocal displacement sensor, we therefore conservatively report a sub-500-nm region.
For water, the minimum measured film thickness was approximately 1~\textmu m.
The larger sheet footprint and stronger gas-assisted thinning observed for acetonitrile are consistent with its lower viscosity and lower surface tension compared with water, in line with the sensitivity of gas-focused liquid structures to liquid properties~\cite{zahoor2020numerical,belsak2021computational}.
The lower viscosity allows the liquid sheet to redistribute more readily under the momentum transferred from the assisting gas jet, while the lower surface tension reduces the capillary restoring force that resists formation of an ultrathin film~\cite{bush2003influence}.
Because the present measurements resolve the liquid-surface profile but not the local gas--liquid stress distribution or in-plane velocity field, we describe the effect phenomenologically as gas-assisted shaping and thinning.
\par
The elongated thinned channel marks the path over which the assisting gas jet reshapes the liquid sheet.
Figures~\ref{fig8}(c) and \ref{fig8}(d) show X- and Y-direction lineouts extracted from the acetonitrile map in Fig.~\ref{fig8}(a), while Figs.~\ref{fig8}(e) and \ref{fig8}(f) show the corresponding lineouts from the water map in Fig.~\ref{fig8}(b).
The small modulations in the acetonitrile longitudinal lineout in Fig.~\ref{fig8}(d) are consistent with the residual scan-line artifact described in Sec.~\ref{sec:setup}.
For acetonitrile, the gas jet produces a long, narrow ultrathin channel along its propagation path, extending over a region of approximately $3~\mathrm{mm}$ in length and $0.5~\mathrm{mm}$ in width.
For water, the minimum measured film thickness is reduced to approximately 1 \textmu m, but the surrounding region thickens more rapidly than in the acetonitrile case.
In this water geometry, a grazing-incidence X-ray beam would intersect the hydraulic jump displaced by the air jet, where the liquid thickness reaches approximately 26 \textmu m.
This comparison shows that gas-assisted thinning can produce micrometer-scale water films and sub-500-nm acetonitrile films, but that the usable X-ray footprint must be chosen to avoid both the hydraulic-jump rim and visible ejected droplets.

\section{Implications for time-resolved grazing-incidence X-ray scattering}
\label{sec:giwaxs_implications}

The gas-assisted acetonitrile sheet in Fig.~\ref{fig8}(a) is the configuration most directly relevant to future time-resolved grazing-incidence X-ray scattering measurements.
The key experimental requirements are a continuous ultrathin region that is longer than the projected X-ray footprint, a liquid path length short enough to suppress bulk-liquid scattering and pump--probe temporal smearing, and stable operation without visible droplet emission near the illuminated region.

For a vertically focused X-ray beam of height $w_z$ incident at a grazing angle $\alpha_X$, the projected footprint is
\begin{equation}
L_{\mathrm{fp}}=\frac{w_z}{\sin\alpha_X}.
\end{equation}
As a representative hard-X-ray XFEL case, consider a 20~keV beam incident near the Ag critical angle, $\alpha_X\simeq0.18^\circ$.
For a $3~\mu\mathrm{m}$ vertical focus, this gives $L_{\mathrm{fp}}\simeq1.0~\mathrm{mm}$.
The acetonitrile channel observed here, approximately $3~\mathrm{mm}$ long and $0.5~\mathrm{mm}$ wide, therefore spans such a grazing-incidence footprint.
The lineout in Fig.~\ref{fig8}(d) also shows that the thickness increases only gradually along this channel; between 1 and 4~mm along the gas-thinned path, the average surface slope is approximately $0.04^\circ$, smaller than the grazing-incidence angle considered here.
This allows the X-ray footprint to be aligned along the flattest portion of the mapped sheet.

The same geometry is relevant for pump--probe temporal resolution.
The wavelength-dependent optical group-delay mismatch per propagated acetonitrile (MeCN) path is
\begin{equation}
g_{\mathrm{MeCN}}(\lambda_{\mathrm{pump}})
=
\frac{n_g(\lambda_{\mathrm{pump}})-1}{c},
\end{equation}
where $n_g$ is the optical group index and the hard-X-ray group velocity has been approximated as $c$.
Using reported dispersion data for acetonitrile, $g_{\mathrm{MeCN}}$ is approximately $1.1$--$1.6~\mathrm{fs}/\mu\mathrm{m}$ over the pump-wavelength range relevant to 800--266~nm excitation, with the larger values corresponding to the UV~\cite{kozma2005direct,moutzouris2014refractive}.
For an uncompensated pump-probe crossing geometry, a conservative estimate of the liquid-path contribution is
\begin{equation}
\Delta t_{\mathrm{liq}}^{\mathrm{uncorr}}
\simeq
g_{\mathrm{MeCN}}(\lambda_{\mathrm{pump}})
\frac{h}{\sin\alpha_X}.
\end{equation}
For the 20~keV Ag example above, a 266~nm pump, and a representative acetonitrile thickness of $h=300~\mathrm{nm}$, this gives $\Delta t_{\mathrm{liq}}^{\mathrm{uncorr}}\simeq150~\mathrm{fs}$.
For thicker regions, smaller grazing angles, or larger beam footprints, the footprint-sweep contribution can be mitigated by crossed-beam velocity matching. 
The optical incidence angle or pulse-front geometry can be chosen to match the pump arrival-time gradient to the grazing-incidence X-ray footprint, following the same velocity-mismatch principle used in crossed-beam ultrafast diffraction experiments~\cite{williamson1993ultrafast}.
For co-propagating optical and X-ray projections along the footprint, the first-order velocity-matching condition is
\begin{equation}
n_g(\lambda_{\mathrm{pump}})\cos\alpha_L
=
\cos\alpha_X,
\end{equation}
where $\alpha_L$ is the internal optical grazing angle measured from the liquid surface.
For 266~nm excitation in acetonitrile at $\alpha_X=0.18^\circ$, this condition gives $\alpha_L\simeq48^\circ$ inside the liquid, corresponding by Snell's law to approximately $22^\circ$ grazing incidence from air.
In the ideal ray-optics limit, this geometry cancels the first-order arrival-time gradient across the X-ray footprint.
The remaining depth-dependent contribution is
\begin{equation}
\Delta t_z
\simeq
\frac{\left(n_g\sin\alpha_L-\sin\alpha_X\right)h}{c}.
\end{equation}
For 266~nm excitation in acetonitrile at the velocity-matched angle, this corresponds to approximately
$3.7~\mathrm{fs}/\mu\mathrm{m}$ of physical film thickness.
The residual thickness contribution is therefore only $\sim1.1~\mathrm{fs}$ for a representative
$300~\mathrm{nm}$ region, and remains below $\sim7.5~\mathrm{fs}$ even for a much thicker
$2~\mu\mathrm{m}$ region.
Thus, after velocity matching, the dominant timing constraint is no longer the grazing-incidence footprint sweep, instead, the effective response is set by the pump and X-ray pulse durations, timing jitter, pulse-front quality, beam sizes, and the measured thickness distribution $h(x,y)$ over the selected region of interest.

The water result in Fig.~\ref{fig8}(b) is useful as a micrometer-scale operating point but is more restrictive for ultrafast grazing-incidence measurements.
A $1~\mu\mathrm{m}$ water film at $\alpha_X=0.1^\circ$ gives a larger uncorrected liquid-path contribution than the sub-500-nm acetonitrile geometry, and the measured water geometry provides a shorter thin region before the displaced hydraulic jump.
For water operation, the practical strategies are to use a smaller vertical X-ray beam, a slightly larger grazing angle when compatible with the desired surface sensitivity, and, most importantly, a truncated substrate that keeps the hydraulic jump outside the footprint.
The operating conditions should also avoid visible ejected droplets near the interaction region, as such droplets may introduce uncontrolled scattering background.

\section{Conclusions}

We have developed a method for generating free-flowing, surface-attached ultrathin liquid sheets suitable for ultrafast X-ray scattering at solid–liquid interfaces.
By directing a liquid jet onto a solid surface, we form stable surface-attached sheets with controllable film thicknesses.
We introduced a gas-assisted dual-capillary configuration that substantially reduces film thickness to near the instrumental limit of 250 nm and extends the ultrathin region to millimeter-scale lengths.
\par
The system allows precise control over the liquid flow velocity, incidence angle, and capillary diameter, enabling systematic optimization of film geometry and uniformity.
The confocal displacement sensor provides three-dimensional reconstruction of the sheet profile with submicrometer vertical precision in the film thickness, enabling quantitative assessment of both sheet morphology and temporal stability.
We characterized the dependence of sheet morphology on incidence angle, jet velocity, and jet diameter, and found that, under the tested conditions, stable sheets are maintained below approximately 30\textdegree{} incidence angle and 60~m/s jet velocity.
\par

This work establishes a robust approach for preparing surface-attached liquid sheets with tunable geometries and thicknesses, providing an enabling platform for time-resolved hard-X-ray scattering studies at solid-liquid interfaces.
With appropriate adaptations of the substrate and sample environment, the same concept may also be useful for soft-X-ray spectroscopy of continuously refreshed thin liquids and for reflection or grazing-incidence ultrafast electron diffraction from liquid-covered surfaces.

\begin{acknowledgments}
This work was supported by the Department of Energy, Laboratory Directed Research and Development program at SLAC National Accelerator Laboratory, under Contract No. DE-AC02-76SF00515.
\end{acknowledgments}

\section*{Data Availability}
The data that support the findings of this study, including the processed height maps and lineout data used to generate the figures, are available from the corresponding author upon reasonable request.

\bibliographystyle{aipnum4-2}
\bibliography{aipsamp}

\end{document}